\documentclass{elsart}
\usepackage{natbib}
\usepackage{graphicx}

\def\plotone#1{\centering \leavevmode
 \includegraphics[width=.95\columnwidth]{#1}}

\newcommand{\nw}{\mbox{nW/m$^2$/sr}}
\newcommand{\um}{\mbox{$\mu$m}}
\newcommand{\nodata}{\mbox{--}}
\begin{document}
\runauthor{Wright}
\begin{frontmatter}
\title{COBE\thanksref{COBE} Observations of the Cosmic Infrared Background}
\thanks[COBE]{The COBE datasets were developed by the NASA Goddard Space 
Flight Center under the guidance
of the COBE Science Working Group and were provided by the NSSDC.}
\author{E. L. Wright}
\address{UCLA Astronomy, PO Box 951562, Los Angeles CA 90095-1562, USA}
\begin{abstract}
The  Diffuse InfraRed Background Experiment (DIRBE) on the COsmic
Background Explorer (COBE) measured the total infrared signal seen from space
at a distance of 1 astronomical unit from the Sun.  Using time variations
as the Earth orbits the Sun, it is possible to remove most of the
foreground signal produced by the interplanetary dust cloud [zodiacal light].
By correlating the DIRBE signal with the column density of atomic hydrogen
measured using the 21 cm line, it is possible to remove most of the
foreground signal produced by interstellar dust, although one must still
be concerned by dust associated with H$_2$ (molecular gas) and H~II (the
warm ionized medium).  DIRBE was not able to determine the Cosmic
InfraRed Background (CIRB) in the 5-60 \um\ wavelength range, but did
detect both a far infrared background and a near infrared background.
The far infrared background has an integrated intensity of about 
34 \nw, while the near infrared and optical extragalactic background
has about 59 \nw.
The Far InfraRed Absolute Spectrophotometer (FIRAS) on COBE has been
used to constrain the long wavelength tail of the far infrared background
but a wide range of intensities at 850 \um\ are compatible with the FIRAS
data.  Thus the fraction of the CIRB produced by SCUBA sources has large
uncertainties in both the numerator and the denominator.
\end{abstract}
\end{frontmatter}

\section{Introduction}
The Diffuse InfraRed Background Experiment (DIRBE) on the COsmic
Background Explorer(COBE) satellite was designed to measure the
Cosmic Infrared Background (CIRB).    The results from the DIRBE
team's analysis of the DIRBE data are described in \cite{hauser/etal:1998}.
These depend in a very fundamental way on the model for the zodiacal or
interplanetary dust cloud that was fit to the variation of the DIRBE signal
as a function of solar elongation \cite{kelsall/etal:1998}.  But this zodiacal
model leaves a large residual intensity at 25 \um\ which must be due to
underestimating the zodiacal background, at least at 25 \um.
New models \cite{wright:1998, gorjian/wright/chary:2000} which address
this problem are used here to derive new lower values for the CIRB.

Other approaches \cite{finkbeiner/davis/schlegel:2000} that use only
a small fraction of the DIRBE data to derive a zodiacal model have
given higher values for the CIRB.  These results are unlikely to be true,
but the dominant systematic uncertainty in deriving the CIRB from 
observations taken 1 AU from the Sun remains the uncertainty in fitting
the zodiacal light.

\begin{figure}[t]
\plotone{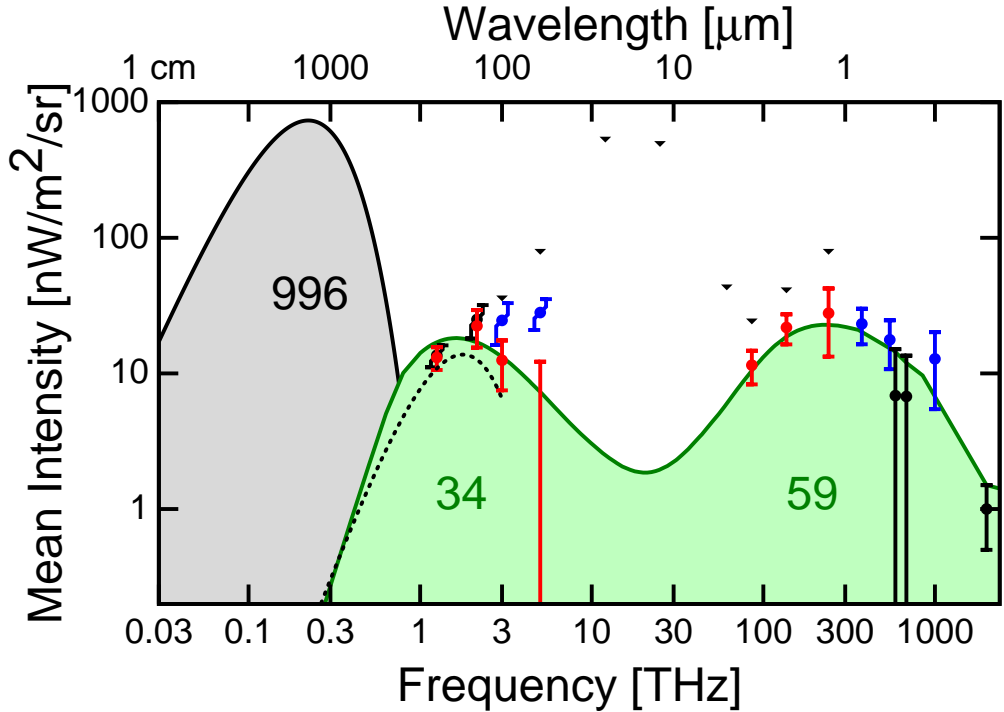}
\caption{The Cosmic Infrared and Optical Background from foreground
subtracted total intensity measurements only.  Black points
and upper limits at $\lambda > 1 \; \um$ are from
\cite{hauser/etal:1998}, while the black points at shorter wavelengths
are from \cite{dube/wicks/wilkinson:1977, toller:1983, 
hurwitz/bowyer/martin:1991}.  The blue points at long wavelengths
are from \cite{finkbeiner/davis/schlegel:2000}, and at short
wavelengths from \cite{bernstein/freedman/madore:2002}.  The numbers
in the bumps indicate the integrated intensity in \nw\ within
each bump.\label{fig:COIBR}}
\end{figure}

\begin{table*}[t]
\caption{DIRBE Derived CIRB Values}
\label{tab:CIRB} 
\vspace{6pt}
\begin{center}
\begin{tabular}{cccc}
\hline 
$\lambda\;[\mu$m] & \multicolumn{3}{c}{$\lambda I_\lambda\;[\nw]$}\\
& This paper & FDS & Hauser {\it et al.} \\
\hline
1.25 & $28\pm 15$      &  \nodata          &  $<75$          \\
2.2  & $22\pm     6$   &  \nodata          &  $<39$          \\
3.5  & $11.5\pm 3.2$   &  \nodata          &  $<23$          \\
60   & $-8\pm 14$      & $28.1 \pm      7$ &  $<75$          \\
100  & $12.5\pm 5$     & $24.6 \pm      8$ &  $<38$          \\
140  & $22 \pm      7$ &  \nodata          &  $25 \pm 6.9$   \\
240  & $13 \pm 2.5$    &  \nodata          &  $13.6 \pm 2.5$ \\
\hline 
\end{tabular}
\end{center}
\end{table*}

\begin{figure}[t]
\plotone{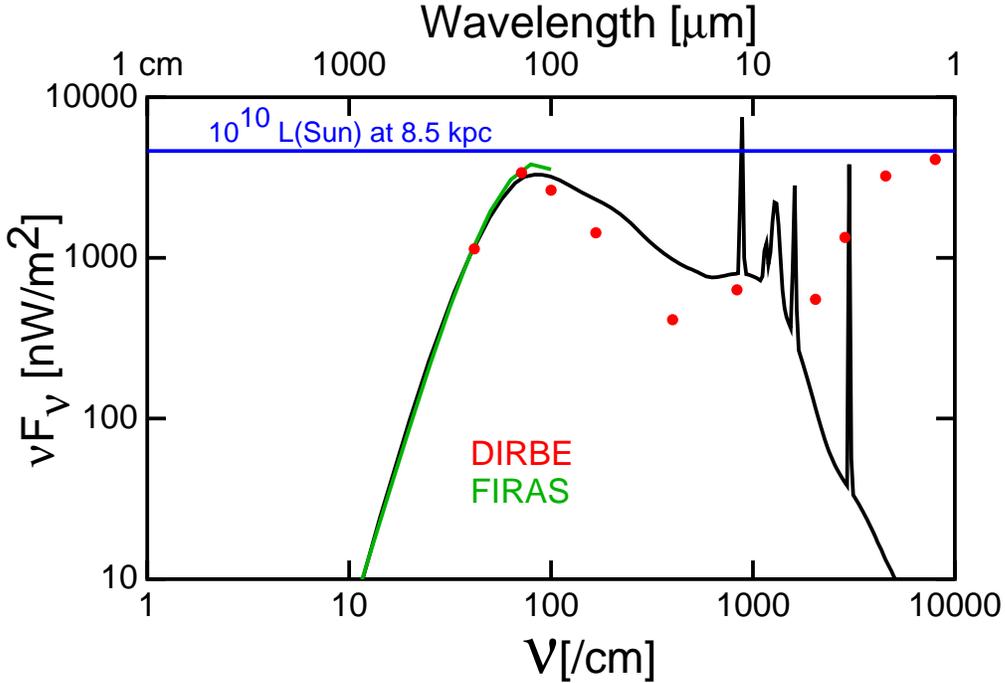}
\caption{Milky Way fluxes measured by FIRAS and DIRBE on COBE.
The black curve is the dust model from \cite{dwek/etal:1998}
for $\log(L/L_\odot) = 10.25$ at 8.5 kpc distance.
\label{fig:MWF}}
\end{figure}

\begin{table*}[t]
\caption{DIRBE Derived Milky Way Fluxes}
\label{tab:MWF} 
\vspace{6pt}
\begin{center}
\begin{tabular}{ccc}
\hline 
$\lambda\;[\mu$m] & $\lambda F_\lambda\;[\mbox{$\mu$W/m$^2$}]$
& $F_\nu\;[\mbox{MJy}]$\\
\hline
1.25 & 4.10 & 1.71 \\
2.2  & 3.23 & 2.37 \\
3.5  & 1.34 & 1.56 \\
4.9  & 0.55 & 0.90 \\
12   & 0.63 & 2.53 \\
25   & 0.41 & 3.43 \\
60   & 1.43 & 28.6 \\
100  & 2.64 & 87.9 \\
140  & 3.39 & 158. \\
240  & 1.14 & 90.9 \\
\hline 
\end{tabular}
\end{center}
\end{table*}

\section{The CIRB}

The DIRBE team zodiacal light model \cite{kelsall/etal:1998} leaves nearly
2 MJy/sr at 25 \um\ in dark regions of the sky.  This is about 6\% of the
zodiacal signal, and gives a reasonable estimate of the uncertainty in the
zodiacal modeling.  This residual intensity corresponds to about
1 photon/cm$^3$/octave.  The lack of a huge $\gamma$-ray absorption at 
10 TeV energy implies that most of this residual intensity is in fact
due to errors in modeling the zodiacal light.  If I add a requirement
that the mean high galactic latitude residual intensity should be zero
to the standard DIRBE zodiacal light modeling, then I get a model 
\cite{gorjian/wright/chary:2000} with a
different geometrical shape for the zodiacal cloud
which gives different estimates for the zodiacal light in
all DIRBE bands, not just the 25 \um\ band.
Using these new estimates for the zodiacal light, the value of the estimated
CIRB at 100 \um\ changes from $20 \pm 5\;\nw$, which was only quoted as an
upper limit in \cite{hauser/etal:1998}, to the lower values given in
Table \ref{tab:CIRB} and shown on Figure \ref{fig:COIBR}.  This change has
a moderate effect on the estimated 140 \um\ CIRB as well, but very little
effect at 240 \um.

Using a statistical argument \cite{wright/reese:2000} or
ground-based observations
\cite{gorjian/wright/chary:2000, wright/johnson:2001}
to remove the galactic foreground stars
allows one to derive estimates instead of
upper limits in the short wavelength bands.  The net result is a CIRB
with nearly twice as much energy in the near infrared and optical bump as in 
the far infrared bump.  The very common statement that the far infrared CIRB 
is larger than the near infrared and optical CIRB is an error caused by using
the {\em lower limits} given by source counts as actual intensities.
Optical work \cite{bernstein/freedman/madore:2002} shows the same effect:
the measured background is about twice the lower limit derived from source
counts.  This is presumably due to the faint fuzzy edges of galaxies being
missed in total flux calculations \cite{wright:2001}.

\section{Milky Way Flux}

One quantity easily derivable from the DIRBE and FIRAS maps is the flux of
the Milky Way.  There is some ambiguity due to the fact that we are located
inside the Milky Way, but the definition of the flux is easy to compute:
\begin{equation}
F_\nu = \int \int I_\nu(l,b) \cos l \cos b d\Omega.
\end{equation}
This is tabulated in Table \ref{tab:MWF} and plotted in Figure \ref{fig:MWF}.
Note that the power in the near infrared is slightly larger than in the far
infrared.  This is significant since the exactly edge-on orientation of
the Milky Way in our sky strongly suppresses the near infrared flux.  Thus
the Milky Way has a much larger luminosity $\nu L_\nu$ in the near infrared
than the far infrared.  Galaxies near $L_*$ such as the Milky Way should
dominate the integrated extragalactic background light, so the fact that
the near infrared bump in the CIRB is larger than the far infrared bump
is easily understood.

\begin{figure}[t]
\plotone{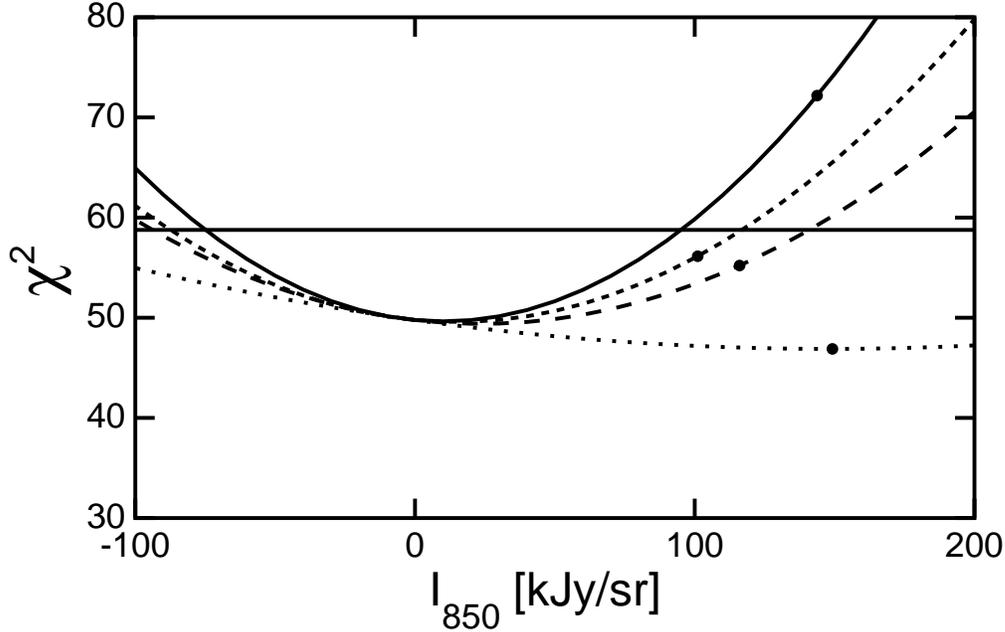}
\caption{$\chi^2$ for the FIRAS low frequency channel data as function
of the 850 \um\ intensity for four different models:
solid \cite{fixsen/etal:1998},  long dashed \cite{lagache/etal:1999},
short dashed is a $\nu^2$ power law \cite{puget/etal:1996}, 
and the dotted curve is a scaled
$\Lambda$CDM Salpeter IMF model \cite{primack/etal:1999} which I have modified
to be consistent with the FIRAS data.  The dot indicates the position
for the published normalization.
The horizontal line is drawn at $\Delta\chi^2 = 9$ units
above the no-CIRB $\chi^2$.
\label{fig:I850}}
\end{figure}

\begin{table*}[t]
\caption{Model CIRB Curve}
\label{tab:I850} 
\vspace{6pt}
\begin{center}
\begin{tabular}{ccccc}
\hline 
$\log(\lambda\;[\mu$m]) & $\log(\lambda I_\lambda\;[\nw])$ &
\quad\quad\quad &
$\log(\lambda\;[\mu$m]) & $\log(\lambda I_\lambda\;[\nw])$ \\
\hline
-7.921 &  -1.921 & &  1.154 &   0.269 \\
-4.523 &  -1.046 & &  1.270 &   0.298 \\
-0.824 &   0.176 & &  1.445 &   0.446 \\
-0.451 &   0.985 & &  1.980 &   1.102 \\
-0.287 &   1.164 & &  2.150 &   1.231 \\
-0.076 &   1.321 & &  2.265 &   1.261 \\
 0.107 &   1.358 & &  2.384 &   1.227 \\
 0.234 &   1.336 & &  2.476 &   1.151 \\
 0.375 &   1.239 & &  2.576 &   1.003 \\
 0.525 &   1.052 & &  2.676 &   0.704 \\
 0.891 &   0.474 & &  3.376 &  -2.122 \\
 1.037 &   0.317 & &  5.276 & -10.250 \\
\hline 
\end{tabular}
\end{center}
\end{table*}

\section{FIRAS Limits on the sub-mm Background}

The final analysis \cite{fixsen/etal:1996} of the FIRAS low frequency
channel data on the Cosmic Microwave Background (CMB) showed a spectrum
with rms residuals from a blackbody of 50 ppm of the peak, or 20 kJy/sr.
How is this result compatible with SCUBA results that claim to
see an integrated intensity from source counts of 86 kJy/sr 
\cite{smail/etal:1998}?
One explanation is that later more extensive observations
\cite{borys/etal:2003} give lower
source counts and $\int S dN = 46$~kJy/sr
for $S > 2$~mJy.
A second explanation is that the FIRAS analysis for CMB distortions
was designed to be insensitive to any foreground that had the spectral
shape of the Milky Way.  Thus the CIRB at 850 \um\ can be quite large
if the CIRB spectrum is similar to the Milky Way spectrum.

Equation (3) from \cite{fixsen/etal:1996} gives this model for small
deviations of the CMB from a blackbody:
\begin{equation}
I_0(\nu) = B_\nu(T_0) + \Delta T {{\partial B_\nu}\over{\partial
T}}
             + G_0 g(\nu) + p {{\partial S_c}\over{\partial p}}
\label{withgal}
\end{equation}
where $p$ describes some distortion.
In order to see how compatible different models for the cosmic sub-mm
background are with the FIRAS low frequency channel data, I compute the
following:
\begin{equation}
\chi^2(p) = \min_{\Delta T, G_0} \left[\sum_{\nu,\nu^\prime}
(I_0(\nu) - p {{\partial S_c}\over{\partial p}})
[C^{-1}]_{\nu,\nu^\prime}
(I_0(\nu^\prime) - p {{\partial S_c}\over{\partial p}})\right],
\end{equation}
with $C_{\nu,\nu^\prime}$ being the covariance matrix,
and let $p$ be the sub-mm background intensity at 850 \um, while
$S_c$ gives the shape of the spectrum.  Thus $\partial S_c/\partial p$
is the spectrum normalized to the 850 \um\ value.
Figure \ref{fig:I850} shows the $\chi^2$ {\it vs.} $I_{850}$ curves for
four different models of the sub-mm background.
The two models with the high values of $I_{850}$ have the best and the worst
$\chi^2$, while the models with lower intensities at 850 \um\ have marginal
$\chi^2$'s.  
The model with the highest $\chi^2$ is the analytic fit
$I_\nu = 1.3 \times 10^{-5} (100\;\um/\lambda)^{0.64} B_\nu(18.5\;\mbox{K})$
from \cite{fixsen/etal:1998}, but this model was designed to fit the peak of
the CIRB near 200 \um, and not the faint low SNR region near 850 \um.
Table \ref{tab:I850} gives the values of $\lambda I_\lambda$ {\it vs.}
$\lambda$ for the model with the best $\chi^2$ in Figure \ref{fig:I850}.
I designed the shape of this model near 850 \um\ to make it compatible
with the FIRAS limits.
This is also the curve shown in Figure \ref{fig:COIBR}.
Clearly more work is needed to determine $I_{850}$, and this
requires a better indicator of the total galactic column density
including all types of matter: H~I, H~II, and H$_2$.

\section{Discussion}

I have presented here results from the DIRBE and FIRAS observations of
the CIRB and the Milky Way.  Both the Milky Way and the CIRB show a
bigger peak in the near infrared than in the far infrared.  The total
intensity is nearly 10\% of the power in the CMB blackbody.  While the
CIRB is probably the integrated light from many unresolved galaxies,
there is a discrepancy between galaxy photometry and the near infrared
and optical CIRB, with the directly measured CIRB being about two times
brighter than the intensity from source counts.  The most probable
explanation of this discrepancy is undercounting the faint outer parts
of galaxies.

\end{document}